\documentclass[showpacs,preprintnumbers,pre]{revtex4}
\usepackage{graphicx}
\usepackage{bm}
\begin{document}

\title{COMPLEX BEHAVIOR IN SIMPLE MODELS OF BIOLOGICAL COEVOLUTION
}

\author{PER ARNE RIKVOLD
}\email{rikvold@scs.fsu.edu}

\affiliation{School of Computational Science, Center for Materials
Research and Technology,\\
Department of Physics, and National High Magnetic Field Laboratory\\
Florida State University, Tallahassee, Florida 32306-4120, USA\\
}

\begin{abstract}
We explore the complex dynamical behavior of simple predator-prey 
models of biological coevolution that account 
for interspecific and intraspecific competition for resources, as well
as adaptive foraging behavior. 
In long kinetic Monte Carlo simulations of these models we 
find quite robust $1/f$-like noise in species diversity and
population sizes, as well as power-law distributions for the
lifetimes of individual species and the durations of quiet
periods of relative evolutionary stasis. 
In one model, based on the Holling Type II functional response,
adaptive foraging produces a metastable low-diversity
phase and a stable high-diversity phase. 
\end{abstract}

\pacs{ 
87.23.Kg, 
05.40.-a, 
05.65.+b 
}

\maketitle

\section{Introduction}
\label{sec:int}

Biological evolution presents a rich array of phenomena that
involve nonlinear interactions between large numbers of units. As a
consequence, problems in evolutionary biology have recently
enjoyed increasing popularity among statistical and computational
physicists.\cite{DROS01} However, many of the models used by
physicists have unrealistic features that prevent the results from
attracting significant attention from biologists. In this
paper we therefore develop and explore individual-based models 
of coevolution in predator-prey systems based on more
realistic population dynamics than some earlier 
models.\cite{HALL02,CHRI02,COLL03,RIKV03,RIKV03A,ZIA04,RIKV05A,SEVI05,SEVI06,RIKV06}

\section{Models}
\label{sec:mod}

Recently the author, together with R.~K.~P.\ Zia, 
introduced a simplified form of the tangled-nature
model of biological macroevolution, which was developed by  
Jensen and collaborators.\cite{HALL02,CHRI02,COLL03} 
In these simplified  
models,\cite{RIKV03,RIKV03A,ZIA04,RIKV05A,SEVI05,SEVI06,RIKV06} 
the reproduction rates in an individual-based population 
dynamics with nonoverlapping generations provide the mechanism for 
selection between several interacting species. New species are
introduced into the community through point
mutations in a haploid, binary ``genome" of
length $L$, as in Eigen's model for molecular 
evolution.\cite{EIGE71,EIGE88}
The potential species are identified by the index $I \in [0,2^L-1]$. 
(Typically, only $\mathcal{N}(t) \ll 2^L$
of these species are present in the community at any one time $t$.) 
At the end of each generation, each individual of species $I$ gives
birth to a fixed number 
$F$ of offspring with probability $P_I$ before dying, or dies
without offspring with probability $(1-P_I)$. Each offspring may
mutate into a different species -- generally with different properties
-- with a small probability $\mu$. Mutation consists in flipping a
randomly chosen bit in the genome. 

\subsection{Simplified tangled-nature models}
\label{sec:tana}

In these models, the reproduction probability for an individual of
species $I$ is given by the nonlinear function 
\begin{equation}
P_I(t) = \frac{1}{1 + \exp[-\Delta_I(R,\{n_J(t)\})]} \;,
\label{eq:PI}
\end{equation}
where $R$ is an external resource that is renewed at the 
same level each generation, and $\{n_J(t)\}$ is the set of
population sizes of all the species resident in the community in
generation $t$. The function $\Delta_I$ is given by  
\begin{equation}
\Delta_I(R,\{n_J(t)\}) = - b_I + \eta_I R /N_{\rm tot}(t) 
+ \sum_J M_{IJ} n_J(t)/N_{\rm tot}(t) - N_{\rm tot}(t)/N_0
\;.
\label{eq:Delta}
\end{equation}
Here $b_I$ is an ``energy cost" of reproduction (always
positive), and $\eta_I$ (positive for primary producers or autotrophs,
and zero for consumers or heterotrophs) is the ability of
individuals of species $I$ to utilize the external resource $R$,
while $N_0$ is an environmental carrying capacity\cite{MURR89} 
(a.k.a.\ Verhulst factor\cite{VERH1838}).
The total population size is $N_{\rm tot}(t) = \sum_J n_J(t)$.
The main feature of this reproduction probability is the random {\it
interaction matrix\/} $\bf M$,\cite{SOLE96} which is constructed at
the beginning of a simulation run, and thereafter kept constant
(quenched randomness). If $M_{IJ}$ is positive and $M_{JI}$
negative, $I$ is a predator and $J$ its prey, and vice versa. If
both matrix elements are positive, the relationship is a 
mutualistic one, while both negative indicate an antagonistic
relationship. 

Two versions of this model were studied in earlier work. In the
first version, which we have called Model A,
there is no external resource or birth cost, and the off-diagonal
elements of $\bf M$ are stochastically independent and uniformly
distributed over $[-1,+1]$, while the diagonal elements are zero. 
This model evolves toward mutualistic communities, in which all 
species are connected by mutually 
positive interactions.\cite{RIKV03,RIKV03A,ZIA04,SEVI05,SEVI06} 

Of greater biological interest is a predator-prey version of the
model, called Model B. In this case a small minority
of the potential species (typically 5\%) are primary producers, while the
rest are consumers. The off-diagonal part of the interaction matrix is
antisymmetric, with the additional restriction that a producer
cannot also prey on a consumer.\cite{RIKV05A,RIKV06} 
In simulations we have taken $b_I$ and the nonzero $\eta_I$ as
independent and uniformly distributed on $(0,+1]$. 
This model generates simple food webs with up to three 
trophic levels.\cite{RIKV05A,RIKV06,RIKV06B}  

Both of these models provide interesting results, which include
intermittent dynamics with power spectral densities (PSDs) of diversities
and population sizes that exhibit $1/f$-like noise, 
as well as power-law distributions for the lifetimes of individual
species and the duration of quiet periods of relative evolutionary stasis.
From a theoretical point of view they
also have the great advantage that the mean-field equation for the
steady-state 
average population sizes in the absence of mutations reduces to a
set of linear (if $N_0 = \infty$) or at most quadratic equations and thus
can easily be solved exactly.\cite{RIKV03,ZIA04,RIKV06,RIKV06B} 
The models thus provide useful benchmarks for more realistic, 
but generally highly nonlinear models. 

In fact, the population dynamics
defined by Eqs.~(\ref{eq:PI}) and~(\ref{eq:Delta}) are not very
realistic. In particular, by summing over positive and negative
terms in $\Delta_I$, the models enable species with little food to
remain near a steady state if they are also not very popular as
prey, or have very low birth cost. 
A more serious problem is the {\it ad-hoc\/} nature of the
normalization by the total population size $N_{\rm tot}(t)$ in the
resource and interaction terms in $\Delta_I$. While this is the source
of the models' analytic solvability, it implies an 
indiscriminate, universal competition without regard to whether or
not two species directly utilize the same resources or share a
common predator. The purpose of the present paper is to develop models
with more realistic population dynamics and explore the
properties of their evolutionary dynamics. 

\subsection{Functional-response model}
\label{sec:holl}

Here we develop a model with more realistic population
dynamics that include competition between different
predators that prey on the same species, as well as a saturation
effect expected to occur for a predator with abundant prey. 
In doing so, we retain from the models discussed above the
important role of the interaction matrix $\bf M$, as well as the
restriction to dynamics with nonoverlapping generations.

We first deal with the competition between predator species
by defining the number of
individuals of $J$ that are available as prey for $I$, corrected for
competition from other predator species, as 
\begin{equation}
\hat{n}_{IJ} = \frac{n_I M_{IJ}}{\sum_L^{{\rm pred}(J)} n_L M_{LJ}} n_J \;,
\label{eq:neff}
\end{equation}
where $\sum_L^{{\rm pred}(J)}$ runs over all $L$ 
such that $M_{LJ} > 0$, i.e.,
over all predators of $J$. Thus, 
$\sum_I^{{\rm pred}(J)} \hat{n}_{IJ} = n_J$, and if $I$ is the only
predator consuming $J$, then $\hat{n}_{IJ} = n_J$. 

Analogously, we define the competition-adjusted
external resources available to a producer species $I$ as 
\begin{equation}
\hat{R}_I = \frac{n_I \eta_I}{\sum_L n_L \eta_L} R \;.
\label{eq:reff}
\end{equation}
As in the case of predators, $\sum_I \hat{R}_I = R$, and a
sole producer species has all of the external resources available to
it: $\hat{R}_I = R$. With these definitions, the total,
competition-adjusted resources available for the sustenance of
species $I$ are 
\begin{equation}
\hat{S}_I = \eta_I \hat{R}_I + \sum_J^{{\rm prey}(I)} M_{IJ} \hat{n}_{IJ} 
\;,
\label{eq:SI}
\end{equation}
where $\sum_J^{{\rm prey}(I)}$ runs over all $J$ such that $M_{IJ} >
0$, i.e., over all prey of $I$. 

A central concept of the model is the {\it functional response\/} of
species $I$ with respect to $J$, $\Phi_{IJ}$.\cite{DROS01B,KREB01} 
This is the rate at which an individual of species $I$ consumes
individuals of $J$. The simplest functional response corresponds to
the Lotka-Volterra model:\cite{MURR89} $\Phi_{IJ} = n_J$ if $M_{IJ} > 0$ 
and 0 otherwise. However, it is reasonable to expect that the
consumption rate should saturate in the presence of very abundant
prey.\cite{KREB01} For ecosystems consisting of a single pair of
predator and prey, or a simple chain reaching from a bottom-level
producer through intermediate species to a top predator, the most
common forms of functional response are due to Holling.\cite{KREB01}
For more complicated, interconnected food webs, a number of
functional forms have been proposed in the recent 
literature,\cite{DROS01B,SKAL01,KUAN02,DROS04,MART06} 
but there is as yet no agreement about a standard form.
Here we choose a ratio-dependent\cite{ABRA94,RESI95} 
Holling Type II form,\cite{KREB01}
\begin{equation}
\Phi_{IJ} = \frac{M_{IJ} \hat{n}_{IJ}}{\lambda \hat{S}_I + n_I} \;,
\label{eq:PhiIJ}
\end{equation}
where $\lambda \in (0,1]$ is the metabolic efficiency of converting
prey biomass to predator offspring. 
Analogously, the functional response of a producer species 
$I$ toward the external resource $R$ is 
\begin{equation}
\Phi_{IR} = \frac{\eta_I \hat{R}_{I}}{\lambda \hat{S}_I + n_I} \;.
\label{eq:PhiIR}
\end{equation}
In both cases, if $\lambda \hat{S}_I \ll n_I$, then the consumption
rate equals the resource ($M_{IJ} \hat{n}_{IJ}$ or 
$\eta_I \hat{R}_{I}$) divided by the number of individuals of $I$,
thus expressing intraspecific competition for scarce
resources. In the opposite limit, $\lambda \hat{S}_I \gg n_I$, 
the consumption rate is proportional to the ratio of the specific,
competition-adjusted resource to the competition-adjusted total
available sustenance, $\hat{S}_I$. The total consumption rate for
an individual of $I$ is therefore 
\begin{equation}
C_I = \Phi_{IR} + \sum_J^{{\rm prey}(I)} \Phi_{IJ} 
= \frac{\hat{S}_I}{\lambda
\hat{S}_I + n_I}
=
\left\{
\begin{array}{lll}
\hat{S}_I/n_I & \mbox{for} & \lambda \hat{S}_I \ll n_I \nonumber\\
1/\lambda     & \mbox{for} & \lambda \hat{S}_I \gg n_I 
\end{array}
\right.
\;.
\label{eq:CI}
\end{equation}
The birth probability 
is assumed to be proportional to the consumption rate, 
\begin{equation}
B_I = \lambda C_I \in [0,+1] \;,
\label{eq:BI}
\end{equation}
while the probability that an individual of $I$ 
avoids death by predation until attempting to reproduce is 
\begin{equation}
A_I = 1 - \sum_J^{{\rm pred}(I)} \Phi_{JI} \frac{n_J}{n_I} \;.
\label{eq:AI}
\end{equation}
The total reproduction probability for an individual of species $I$
in this model is thus $P_I(t) = A_I(t) B_I(t)$. 

\section{Numerical Results for the Functional-response Model}
\label{sec:Sim1}

We simulated the functional-response model over
$2^{24} = 16\,777\,216$ generations (plus $2^{20}$
generations ``warm-up") for the
following parameters: genome length $L=21$ 
($2^{21} = 2\,097\,152$ potential
species), external resource $R=16\,000$, fecundity $F=2$,  
mutation rate $\mu = 10^{-3}$, proportion of
producers $c_{\rm prod} =0.05$, interaction matrix $\bf M$ with
connectance $C = 0.1$ and nonzero elements with a symmetric, 
triangular distribution over $[-1,+1]$, and $\lambda = 1.0$. 
We ran five independent runs, each starting from 100 individuals
of a single, randomly chosen producer species. 

\subsection{Time series}
\label{sec:timser}

\begin{figure}[t]
\begin{center}
\vspace*{0.3truecm}
\includegraphics[angle=0,width=.47\textwidth]{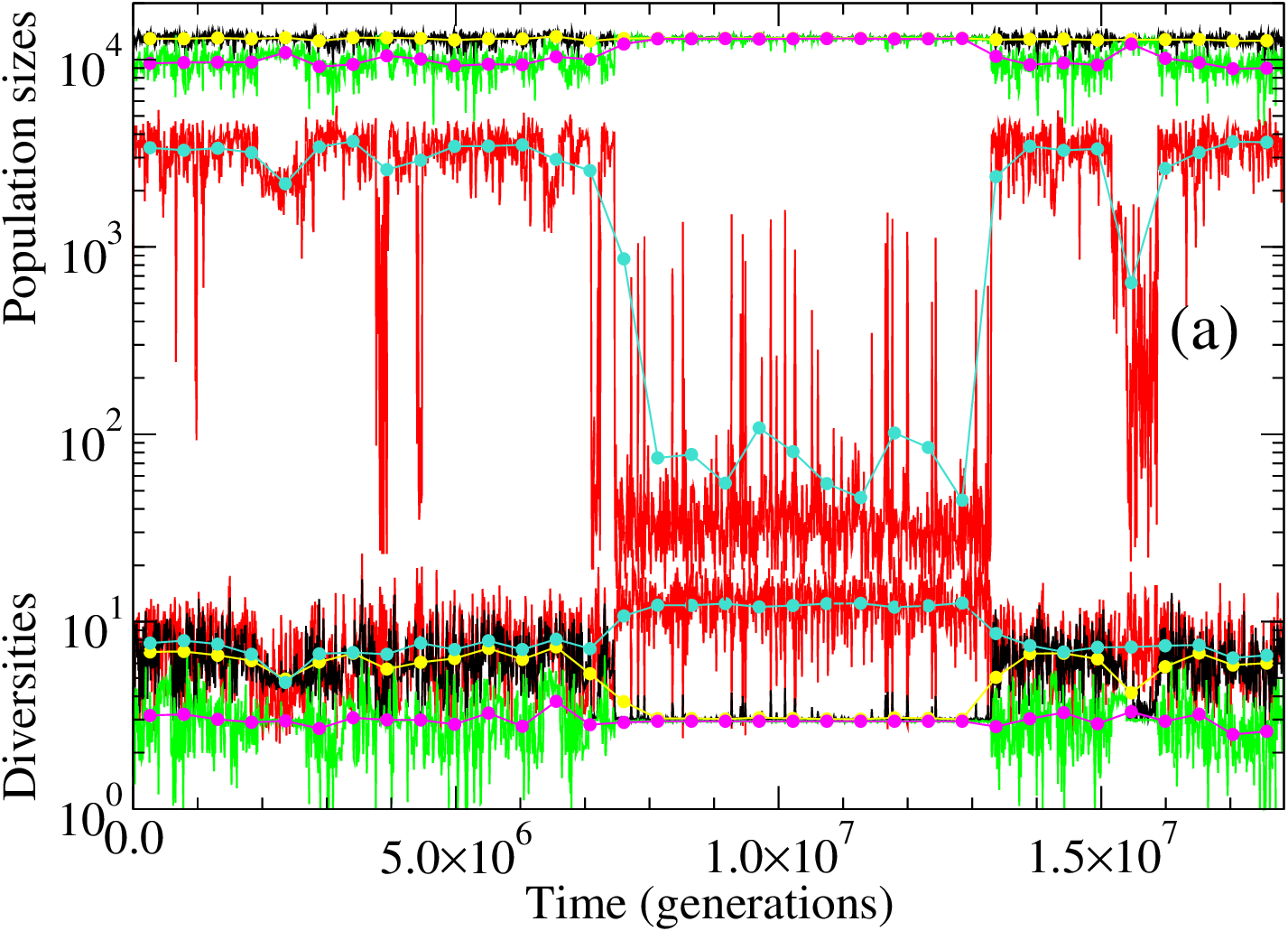}
\hspace{0.5truecm}
\includegraphics[angle=0,width=.47\textwidth]{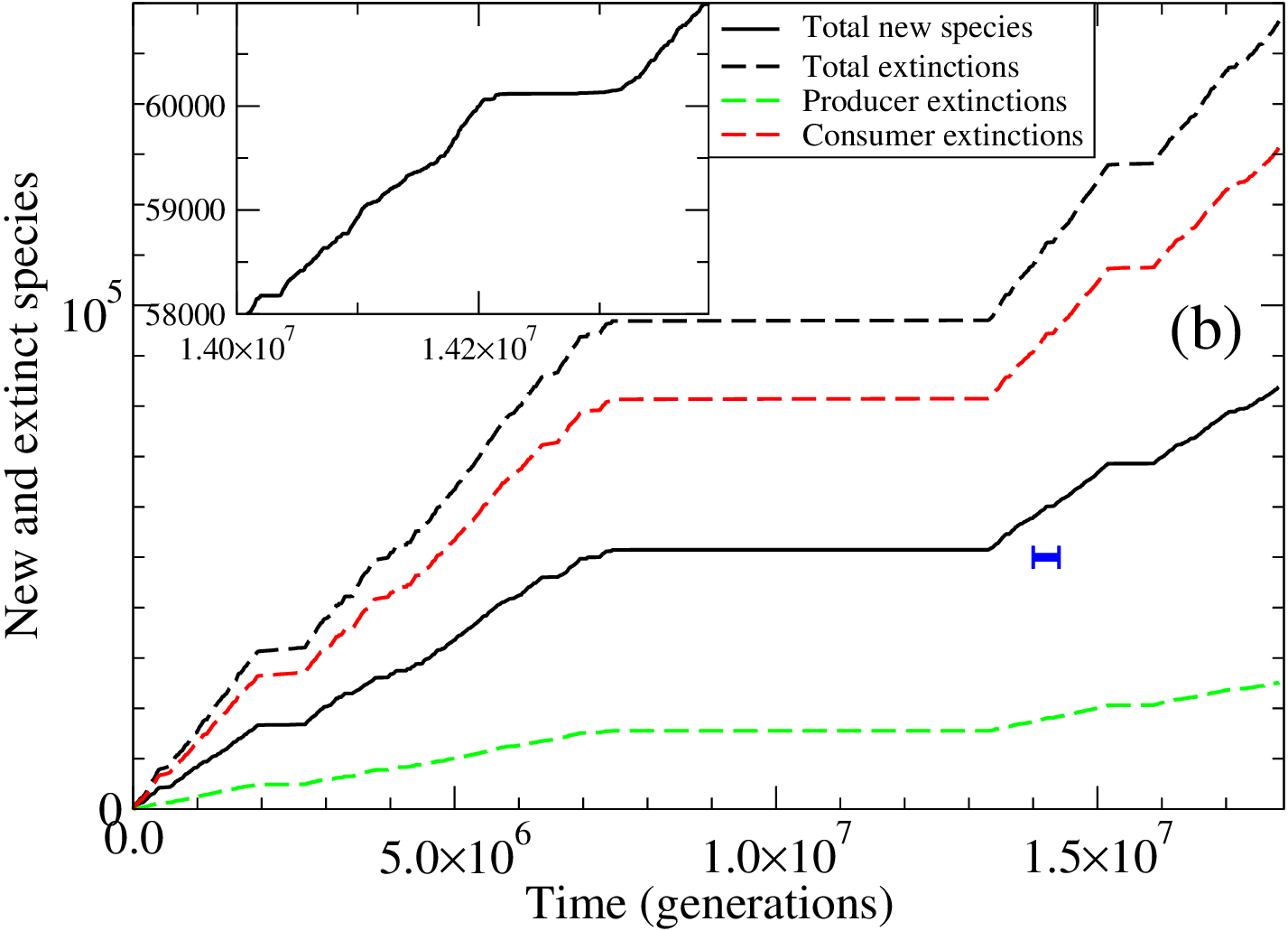}
\end{center}
\caption[]{
(Color online.)
{\bf (a)} 
Time series of diversities (measured in species, lower curves) 
and population sizes (measured in individuals, upper curves) for
one specific simulation run. The
strongly fluctuating curves in the background are sampled every 
8192 generations, while the smooth curves with data points in
contrasting colors that are overlaid 
in the foreground are running averages over
524\,288 generations. Black with light gray (yellow online) 
overlay: all species.  
Light gray with dark gray overlay (green and magenta online): producers. 
Dark gray with light gray overlay (red and cyan online): consumers. 
{\bf (b)} 
Time series of the 
accumulation of new and extinct species in the same simulation run
depicted in (a). 
The solid, black curve shows the total number of different
species that have at least once 
attained a population size $n_I > 1000$ by time $t$. The dashed curves
count the total number of species that have gone extinct after 
attaining a maximum population greater than 1000. The black dashed
curve refers to all species, the light gray one (green online) to
producers, and the dark gray one (red online) to consumers. The
ratio of approximately 1.89 between the dashed and full black curves
indicate that major species recur on average about twice
during the evolution. This is an artifact of the finite genome
length. The inset shows the detailed, intermittent structure of
the solid, black curve over $400\,000$ generations. The interval is
indicated by a horizontal bar in the main panel.  
}
\label{fig:timser}
\end{figure}
Time series of diversities (effective numbers of species)
and population sizes 
for one realization are shown in Fig.~\ref{fig:timser}. To
filter out noise from low-population,  
unsuccessful mutations, we define the diversity as the exponential
Shannon-Wiener index.\cite{KREB89} This is the exponential function 
of the information-theoretical entropy of the population distributions,
$D(t) = \exp \left[S \left( \{ n_I(t) \} \right) \right]$, where
\begin{equation}
S\left( \{ n_I(t) \} \right)
=
- \sum_{\{I | \rho_I(t) > 0 \}} \rho_I(t) \ln \rho_I(t)
\label{eq:S}
\end{equation}
with
$\rho_I(t) = n_I(t) / N_{\rm tot}(t)$ for the case of all species,
and analogously for the producers and consumers separately. 

The time series for both diversities and population sizes show
intermittent behavior with quiet periods of varying lengths,
separated by periods of high evolutionary
activity. In this respect, the results
are similar to those seen for Models A and B in earlier 
work.\cite{RIKV03,RIKV03A,RIKV05A,SEVI06,RIKV06} However, 
diverse communities in this model seem to be less stable than those
produced by the linear models. In particular, this model has a
tendency to flip randomly between an active
phase with a diversity near 
ten, and a ``garden of Eden'' phase of one or a few producers with
a very low population of unstable consumers, such as the one seen
around 10 million generations in Fig.~\ref{fig:timser}.

\subsection{Power-spectral densities}
\label{sec:psd}

\begin{figure}[t]
\begin{center}
\vspace*{0.1truecm}
\includegraphics[angle=0,width=.47\textwidth]{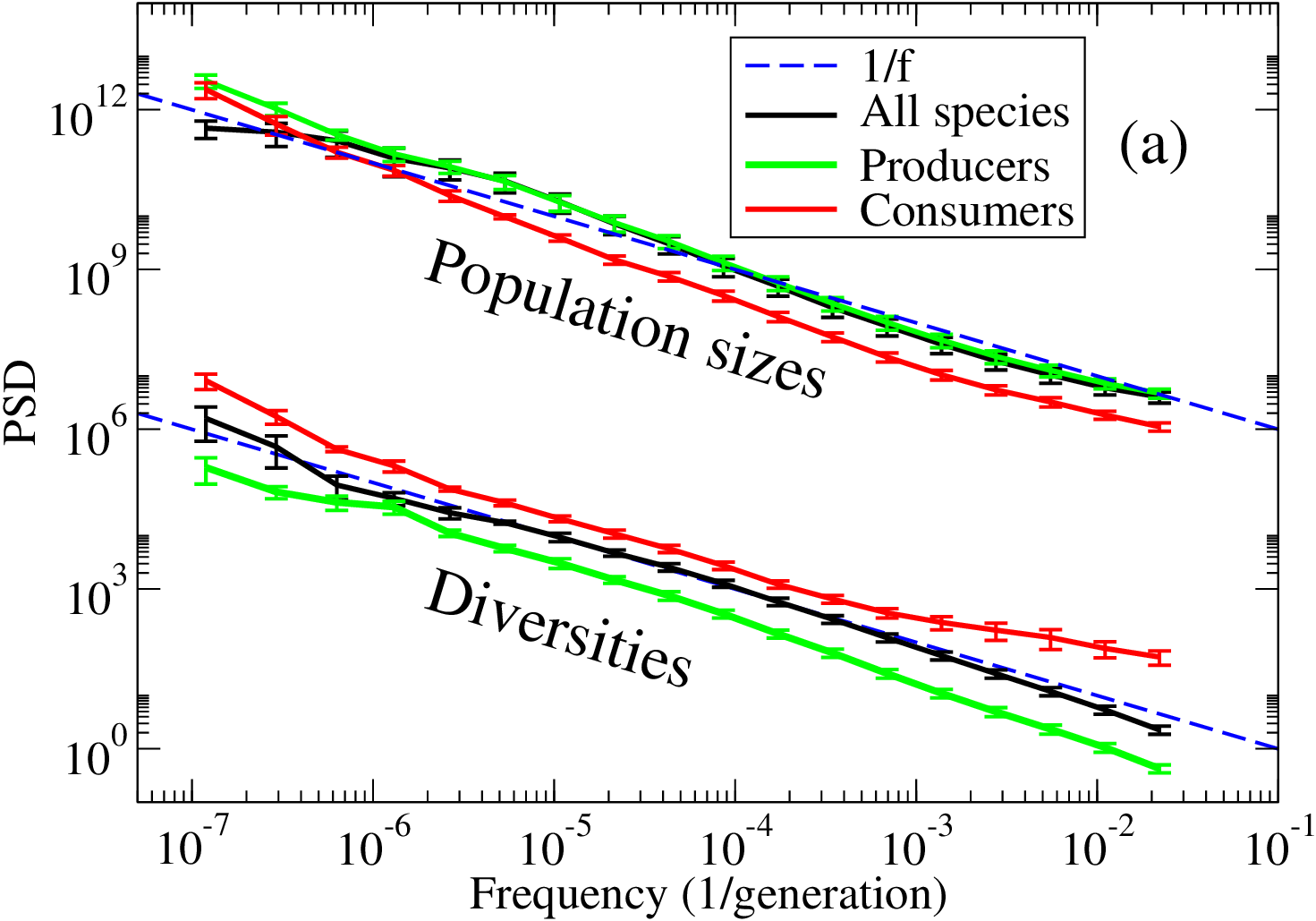}
\hspace{0.5truecm}
\includegraphics[angle=0,width=.47\textwidth]{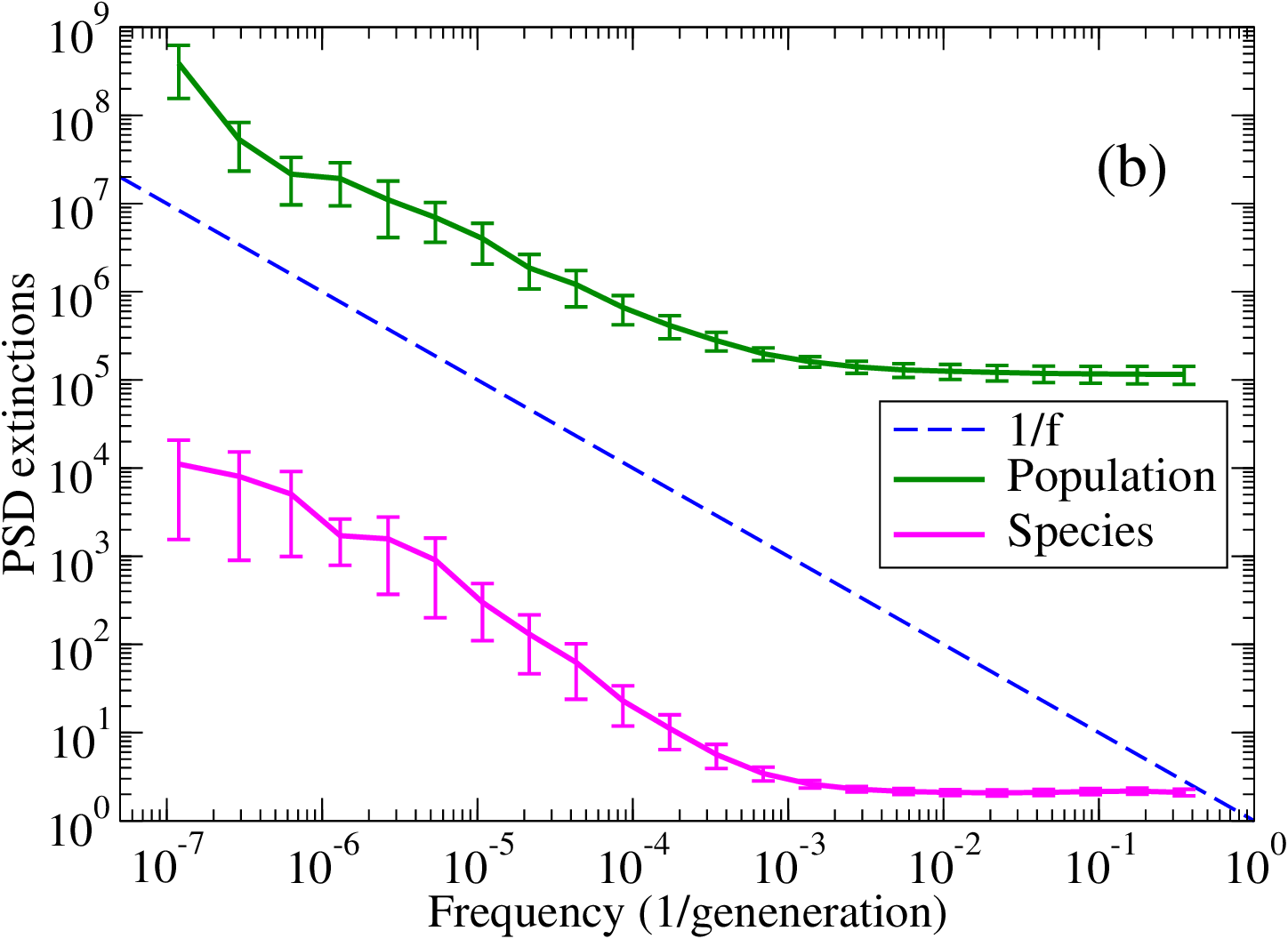}
\end{center}
\caption[]{
(Color online.)
{\bf (a)}
PSDs for the diversities and population sizes, each recorded
separately for all species and for producers and consumers. 
The time series were sampled every 16 generations. 
{\bf (b)}
PSDs for the extinction activity. 
In both parts of the figure, the results are averaged over five
independent simulation runs. 
See discussion in the text. 
}
\label{fig:psd}
\end{figure}
A common method to obtain information about the intensity of
fluctuations in a time series at different time scales is the
power-spectral density (squared Fourier transform), or PSD. PSDs are
presented in Fig.~\ref{fig:psd} for the diversity fluctuations and the
fluctuations in the population sizes (Fig.~\ref{fig:psd}(a))
and the intensity of extinction events (Fig.~\ref{fig:psd}(b)). 
The former two are shown for the total population, as well as 
separately for the
producers and consumers. All three are similar.
Extinction events are recorded as
the number of species that have attained a population size greater
than one, which go extinct in generation $t$ (marked as ``species"
in the figure), while
extinction sizes are calculated by adding the maximum populations
attained by all species that go extinct in generation $t$ (marked
as ``population" in the figure). 
The PSDs for all the quantities shown exhibit approximate $1/f$
behavior. For the diversities and population sizes, this power law
extends over more than five decades in time. The extinction
measures, on the other hand, have a large background of white noise
for frequencies above $10^{-3}$ generations$^{-1}$, 
probably due to the high rate of  
extinction of unsuccessful mutants. For lower frequencies, however,
the behavior is consistent with $1/f$ noise within the limited
accuracy of our results. 

\subsection{Species lifetimes and durations of quiet periods}
\label{sec:times}

\begin{figure}[t]
\begin{center}
\vspace*{1.1truecm}
\includegraphics[angle=0,width=.47\textwidth]{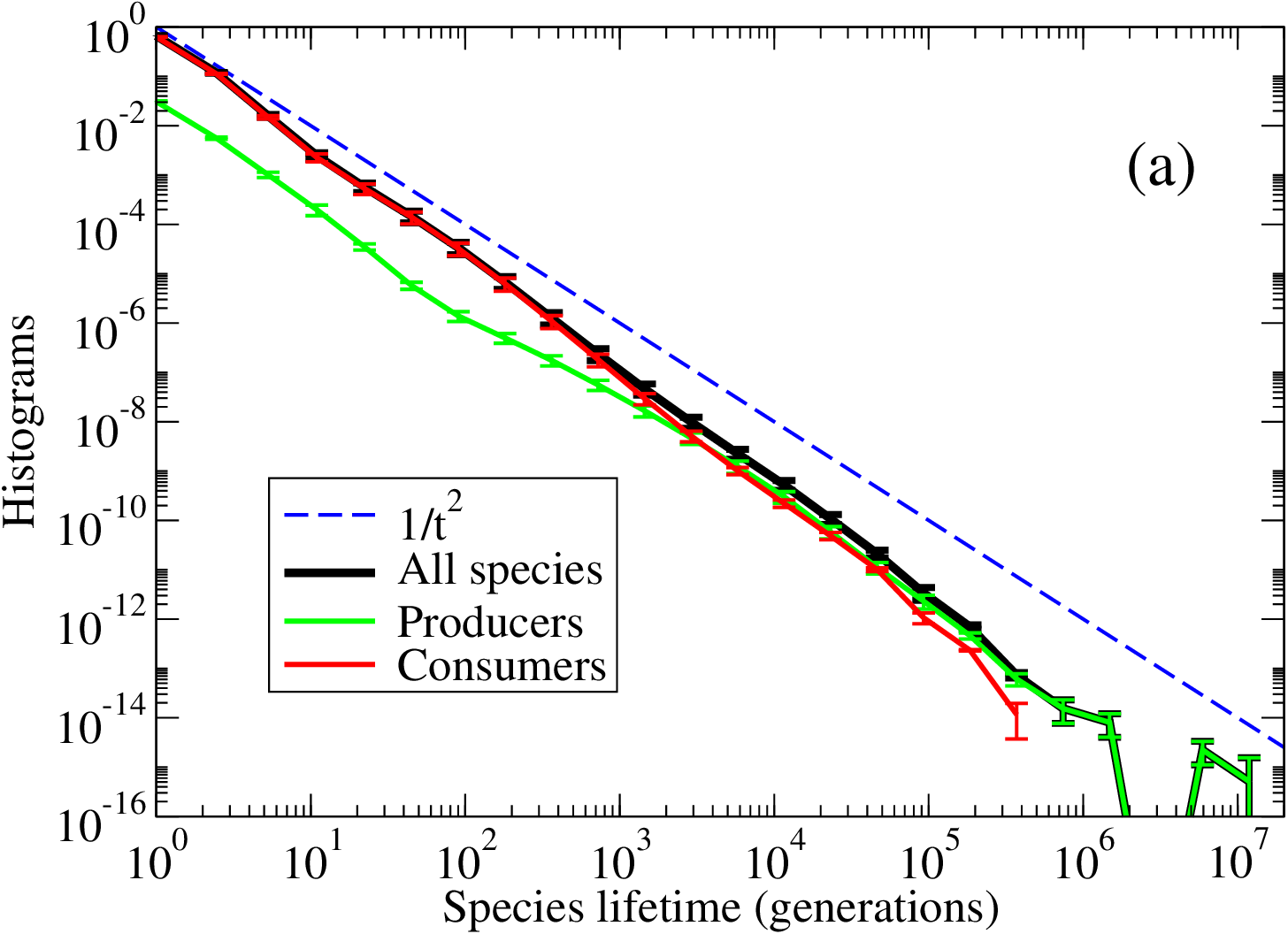}
\hspace{0.5truecm}
\includegraphics[angle=0,width=.47\textwidth]{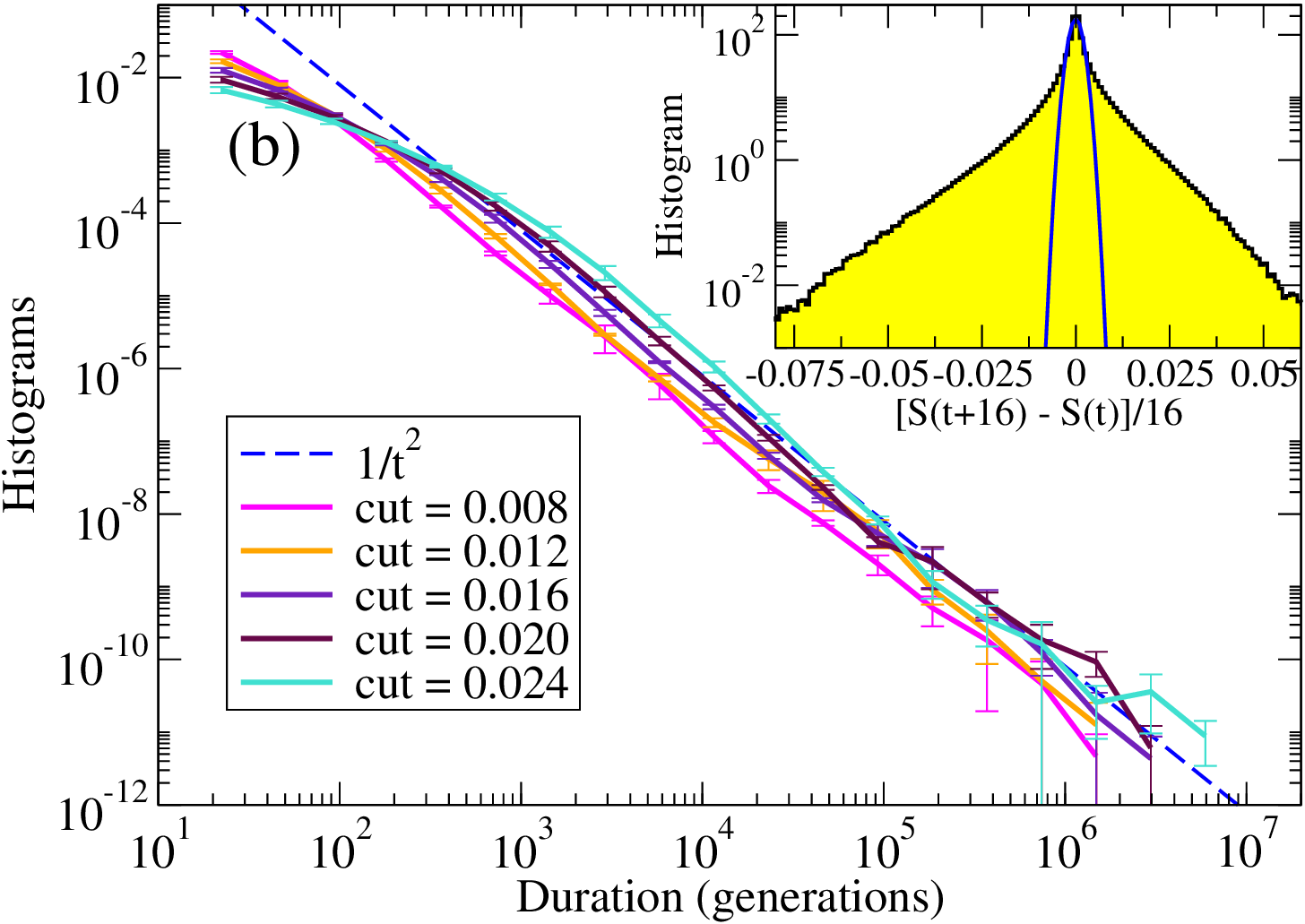}
\end{center}
\caption[]{
(Color online.)
{\bf (a)}
Histograms of species lifetimes, shown for all species, as well as
separately for producers and consumers. 
{\bf (b)}
Histograms of the durations of evolutionarily quiet periods,
defined as the times that $|{\rm d}S/{\rm d}t|$ (averaged over 16 generations)
falls continuously below some cutoff. The inset is a histogram of 
${\rm d}S/{\rm d}t$, showing a Gaussian center with approximately exponential
wings. The parabola in the foreground is a Gaussian fit to this
central peak. The cutoff values for the main figure, 
between 0.008 and 0.024, were chosen on the basis of this distribution. 
The data in both parts of the figure
are averaged over five independent simulation runs. 
}
\label{fig:time}
\end{figure}
The evolutionary dynamics can also be characterized by histograms
of characteristic time intervals, such as the time from creation till
extinction of a species (species lifetimes) or the time intervals
during which  
some indicator of evolutionary activity remains continuously below 
a chosen cutoff (duration of evolutionarily quiet periods). 
Histograms of species lifetimes are shown in Fig.~\ref{fig:time}(a). 
As our indicator of evolutionary activity we use the magnitude of the
logarithmic derivative of the diversity, $|{\rm d}S/{\rm d}t|$, 
and histograms for the resulting durations of
quiet periods, calculated with different cutoffs, are shown in 
Fig.~\ref{fig:time}(b). Both quantities display approximate
power-law behavior with an exponent near $-2$, consistent with the
$1/f$ behavior observed in the PSDs.\cite{RIKV03,PROC83} 
It is interesting to note that the distributions for these two
quantities for this model have approximately the same exponent.
This is consistent with the previously studied, mutualistic 
Model A,\cite{RIKV03,RIKV05A} 
but not with the predator-prey Model B.\cite{RIKV05A,RIKV06,RIKV06B} 
We believe the linking of the power laws for the species lifetimes
and the duration of quiet periods indicate that the communities
formed by the model are relatively fragile, so that all member
species tend to go extinct together in a ``mass extinction." In
contrast, Model B produces simple food webs that are much more
resilient against the loss of a few species, and as a result the
distribution of quiet-period durations decays with an exponent near 
$-1$.\cite{RIKV05A,RIKV06,RIKV06B}

\section{Adaptive Foraging}
\label{sec:adap}

\begin{figure}[t]
\begin{center}
\vspace*{0.2truecm}
\includegraphics[angle=0,width=.47\textwidth]{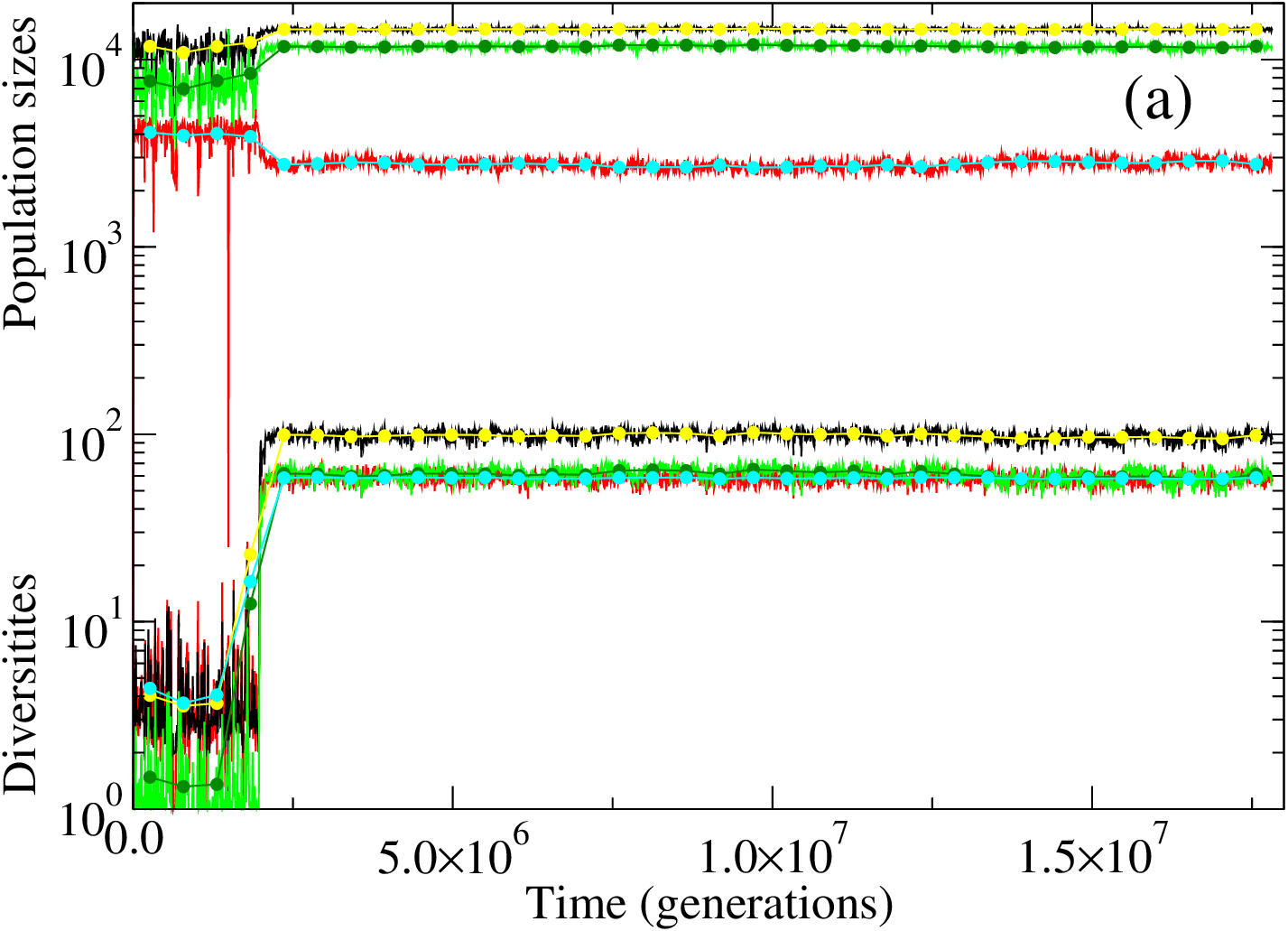}
\hspace{0.5truecm}
\includegraphics[angle=0,width=.47\textwidth]{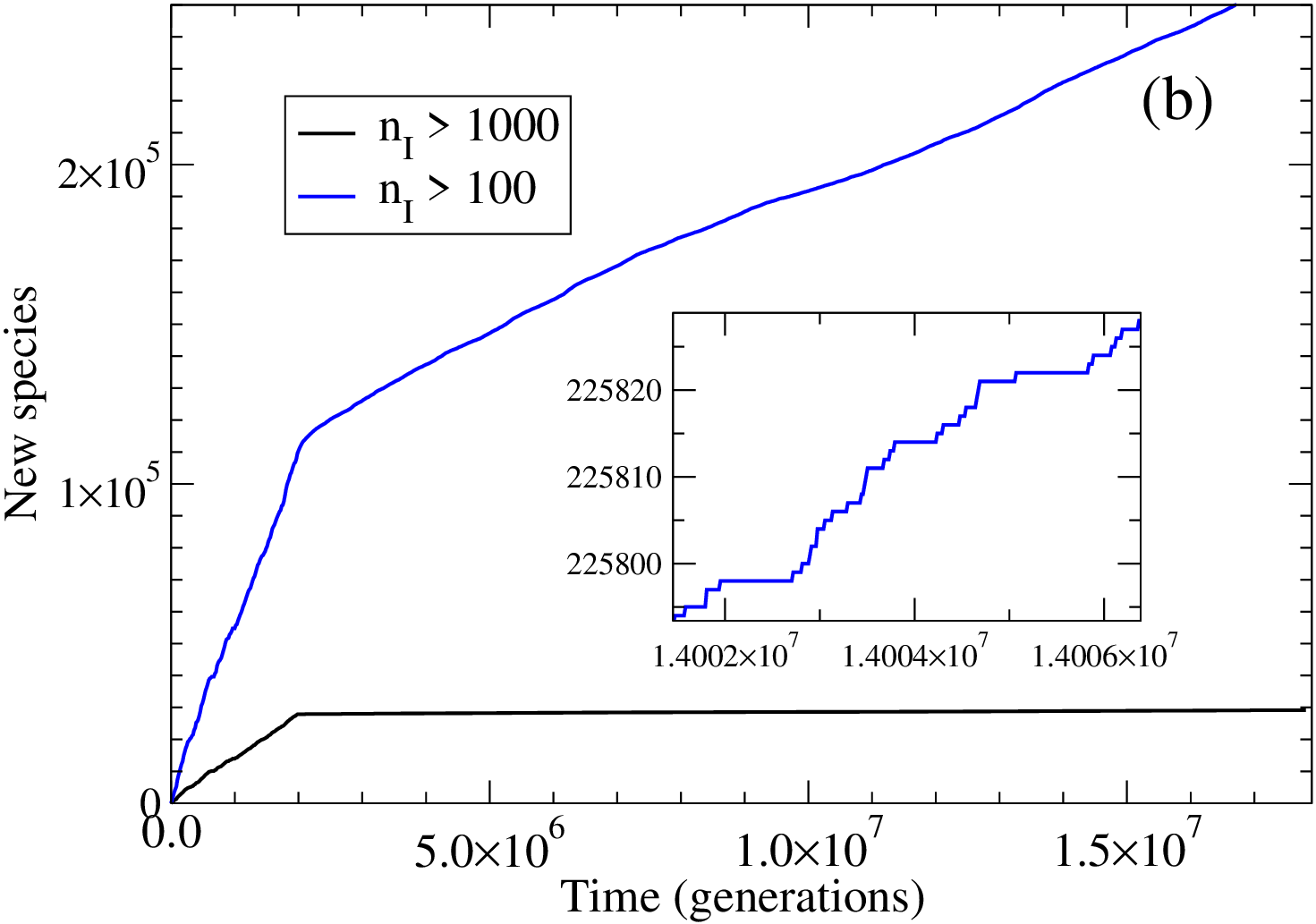}
\end{center}
\caption[]{
(Color online.)
{\bf (a)}
Time series of 
population sizes (upper curves) and diversities (lower curves) for the model
with adaptive foraging. The interpretation of the colors and lines are
the same as in Fig.~\protect\ref{fig:timser}(a). 
{\bf (b)}
Time series of 
the number of new species that have reached a population greater than
1000 (lower curve) and greater than 100 (upper curve). The inset shows
the intermittent structure of the upper curve on a very fine scale of
2000 generations.  See discussion in the text. 
}
\label{fig:adap}
\end{figure}
The model studied above is one in which species forage
indiscriminately over all available resources, with the output only
limited by competition. Also, there is an implication that an
individual's total foraging effort increases proportionally with
the number of species to which it is connected by a positive  
$M_{IJ}$. A more realistic picture would be that an individual's
total foraging effort is constant and can either be divided
equally, or concentrated on richer resources. This is known as
adaptive foraging. While one can 
go to great length devising optimal foraging
strategies,\cite{DROS01B,DROS04} we here
only use a simple scheme, in which individuals of $I$ show a
preference for prey species $J$, based on the interactions and
population sizes (uncorrected for interspecific competition) and given by 
\begin{equation}
g_{IJ} = \frac{M_{IJ}n_J}{\eta_I R + \sum_K^{{\rm prey}(I)} M_{IK} n_K}
\;,
\label{eq:gij}
\end{equation}
and analogously for $R$ by  
\begin{equation}
g_{IR} = \frac{\eta_{I} R}{\eta_I R + \sum_K^{{\rm prey}(I)} M_{IK} n_K}
\;. 
\label{eq:gir}
\end{equation}
The total foraging effort is thus 
$g_{IR} + \sum_J^{{\rm prey}(I)} g_{IJ} = 1$. 
The preference factors are used to modify the reproduction 
probabilities by replacing all occurrences of $M_{IJ}$
by $M_{IJ} g_{IJ}$ and of $\eta_I$ by $\eta_I g_{IR}$ in 
Eqs.~(\ref{eq:neff}--\ref{eq:PhiIR}). 

\begin{figure}[t]
\begin{center}
\vspace*{0.2truecm}
\includegraphics[angle=0,width=.47\textwidth]{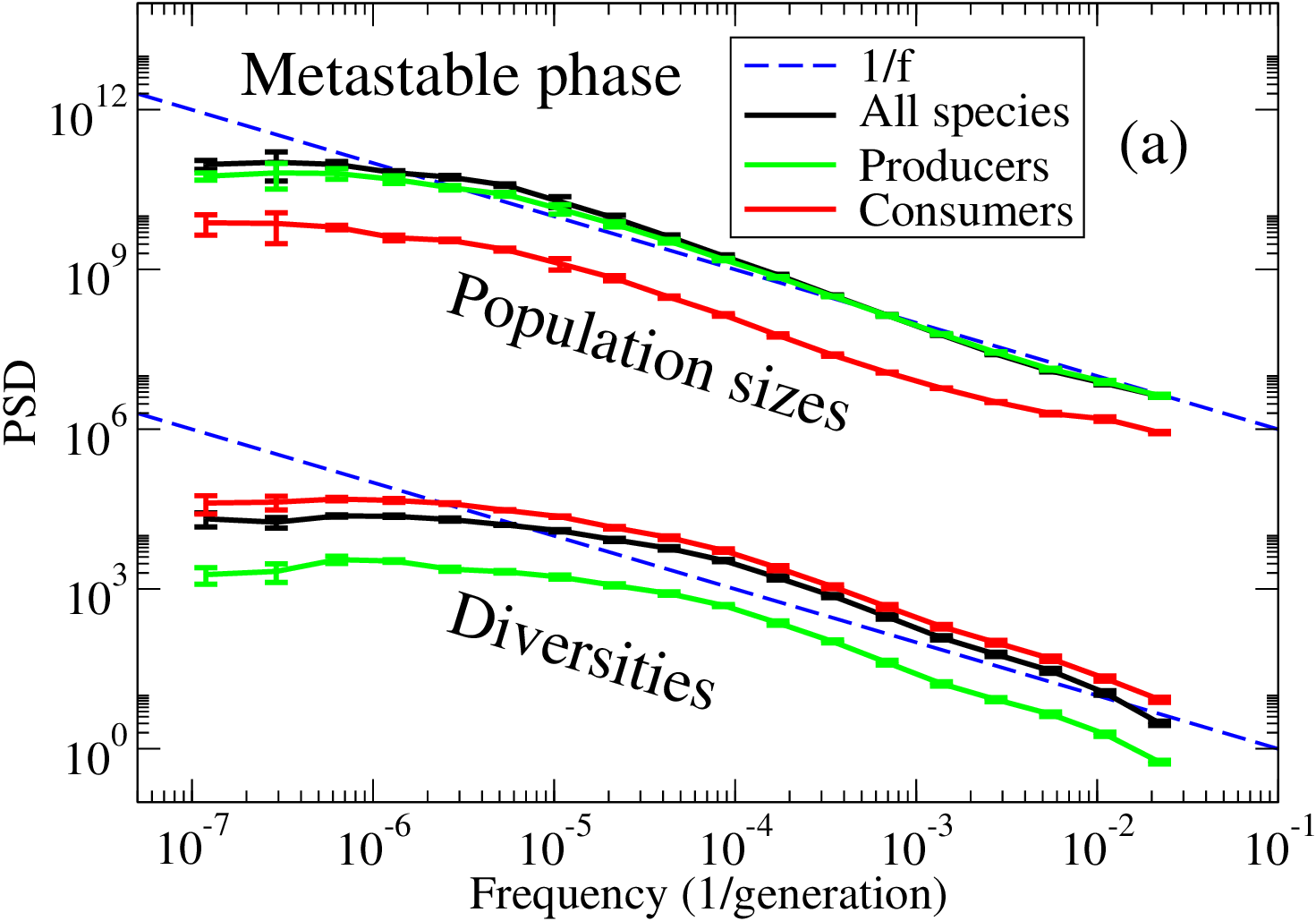}
\hspace{0.5truecm}
\includegraphics[angle=0,width=.47\textwidth]{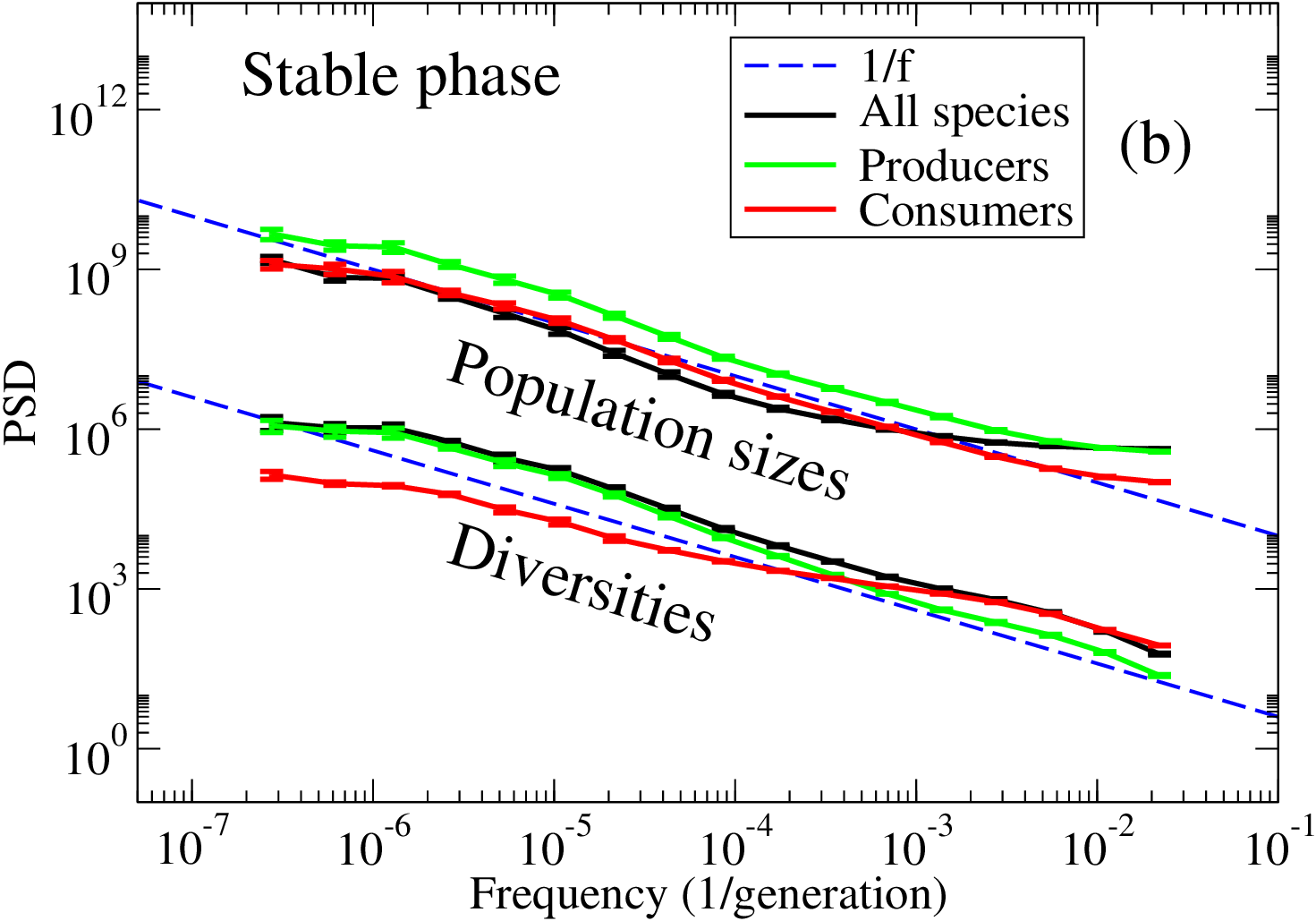}
\end{center}
\caption[]{
(Color online.)
PSDs for the diversities and population sizes in the metastable phase
(averaged over three independent runs)
{\bf (a)}, and the stable phase (averaged over five independent runs)
{\bf (b)} for the model with adaptive foraging. Both show
approximate $1/f$ noise for frequencies above about $10^{-5}$, 
but the PSDs appear to approach constant levels for the lowest
frequencies. 
}
\label{fig:adapPSD}
\end{figure}

The results of implementing the adaptive foraging are quite striking.
The system appears now to have a metastable low-diversity phase 
similar to the active phase of the non-adaptive model, from
which it switches at a random time to an apparently stable
high-diversity phase with much smaller fluctuations. 
As seen in Fig.~\ref{fig:adap}(a), the switchover
is quite abrupt, and Fig.~\ref{fig:adap}(b) shows that it is accompanied by
a sudden reduction in the rate of creation of new species.
As seen in Fig.~\ref{fig:adapPSD}, the PSDs for both the diversities and
population sizes in both phases show approximate $1/f$ noise for
frequencies above $10^{-5}$ generations$^{-1}$. For lower
frequencies, the metastable phase shows no discernible frequency
dependence, while for the stable phase, the frequency dependence continues 
at least another decade. It thus appears that long-time correlations are
not seen beyond $10^5$ generations for the metastable phase, and
probably not beyond about $10^6$ generations for the stable one. 
These observations are 
consistent with species-lifetime distributions for both phases
(not shown), which are quite similar to those for the
non-adaptive model, but typically with cutoffs in the range of $10^5$ to
$10^6$ generations, much shorter than the total simulation times. 

In fact, the system can also escape from the
low-diversity phase to total extinction, which is an absorbing state,
and in some of our simulation runs we avoided this by limiting
$|M_{IJ}|$ to less than 0.9. This restriction does not seem to have any
effect on the dynamics in the high-diversity phase. 
These results are preliminary, and it is possible that the
high-diversity phase corresponds to a mutational meltdown. More research
is clearly needed regarding the effects of adaptive foraging in this
model.

\section{Conclusions}
\label{sec:conc}

In this paper we have shown that very complex and diverse dynamical
behavior results, even from highly over-simplified models of biological
macroevolution. In particular, PSDs that show $1/f$-like noise and
power-law lifetime distributions for species as well as evolutionarily quiet
states are generally seen. This is the case, both in the analytically
tractable, but somewhat unrealistic tangled-nature type models, and in
the nonlinear predator-prey models based on the more realistic Holling Type II
functional response. Particularly intriguing is the appearance of a new,
stable high-diversity phase in the latter type of model when adaptive
foraging behavior is included. Among the many questions about this new
phase that remain to be addressed is the structure of the resulting
community food webs.

\section*{Acknowledgments}

Supported in part by U.S.\ National Science Foundation Grant Nos.\
DMR-0240078 and DMR-0444051 and by Florida State University through the
School of Computational Science, the Center for Materials Research
and Technology, and the National High Magnetic Field Laboratory.




\end{document}